\documentclass[prl,twocolumn,amsmath,showpacs]{revtex4}
\usepackage{graphicx}
\usepackage{amssymb}
\usepackage{citesort}
\usepackage{stmaryrd}

\usepackage{graphicx}

\begin{document}

{\bf Comments on " Controlling Discrete and Continuous Symmetries in Superradiant Phase Transitions with Circuit QED Systems " }

   Recently, the authors in Ref.\cite{comment} presented the $ N=\infty $ solution of the $ U(1)/Z_2 $ Dicke model \cite{gold}.
   Here we point out that
   (1) The authors missed an important transformation relating the two parameter regimes, so their separate discussions on the two regimes is redundant.
   {\bf Most importantly:} (2) Both $ N=\infty $ classical limit and $ 1/N $ quantum fluctuations have been achieved
   in \cite{gold,berryphase}. It is the $ 1/N $  quantum fluctuations which
   lead to non-trivial new  quantum phenomena. In view of only a few $ N=2\sim 9 $ qubits inside a circuit QED microwave cavity,
   they can be tested in near future experiments.
   (3) Several possible experimental implementations  of the $ U(1)/Z_2 $ Dicke model have been proposed before and
       recently experimentally realized.

  {\bf  1.}  The authors in \cite{comment} claimed there are different physics in two different regimes
  $  \Omega_E > \Omega_M$  and  $  \Omega_E < \Omega_M $. In fact, there is a transformation ( see below )
  establishing the relation  $ \Omega_E \leftrightarrow \Omega_M $. So their separate discussions on the two regimes
  are redundant.

  In the Eqn.1 called $ U(1)/Z_2 $ Dicke model in our work \cite{gold}, $ g $ stands for the Rotating Wave  ( RW) term, and $ g^{\prime} $ stands for the Counter-Rotating wave (CRW) term.
  It is straightforward to see that under $ a \rightarrow e^{ i \pi/2 } a, \sigma_{-}  \rightarrow e^{ i \pi/2 } \sigma_{-} $,
  the CRW term $ g^{\prime} \rightarrow -g^{\prime} $. So in \cite{gold}, we only focused on $ 0 < g^{\prime} < g $ case.
  Obviously, $ g^{\prime}=0 $ is the $ U(1) $ symmetry point, any $ g^{\prime} \neq 0 $ breaks the $ U(1) $ down to $ Z_2 $.
  The authors in Ref.\cite{comment} re-wrote the Eqn.1 in \cite{gold}
  as their Eqn.1 with the straightforward relations  $  g=  \Omega_E + \Omega_M,
  g^{\prime}= \Omega_E - \Omega_M$. Under the same transformation, $ \Omega_E \leftrightarrow \Omega_M$, so one only need to focus on
  $ \Omega_E > \Omega_M$ at both $ N=\infty $ and finite $ N $.
  Below Eqn.13 in \cite{gold}, we found the critical strength $ g+ g^{\prime}= g_c= \sqrt{\omega_a \omega_b} $ which implies
  $ \Omega_E= \sqrt{\omega_a \omega_b}/2 $ shown in Fig.1 and 2 in \cite{comment}.

  {\bf 2.} The results shown in Fig.1-5 in \cite{comment} are just the $ N = \infty  $ ( classical ) limit of
    the effective action Eqn.12 at the order $ 1/N $ in \cite{gold}. It holds for any $ g $ and $ g^{\prime} $.
    Because the classical limit is technically trivial and not useful in any practical circuit QED system, it was only briefly mentioned below Eqn.13 in \cite{gold}. For example,  the Fig.5c in \cite{comment} corresponds to $ g^{\prime }=0 $ in \cite{gold}, namely, the $ U(1) $ Dicke model.
    The Goldstone mode  at $ N =\infty $ is just the flat zero mode \cite{gold,berryphase}, the amplitude model is nothing but the Higgs mode.
    Both modes are explicitly stressed in the title of \cite{gold}.

    The most important value of Eqn.12 is that it can be used to calculate quantum fluctuations at $ 1/N $ at any $ g $ and $ g^{\prime} $.
    In \cite{gold,berryphase}, we computed the quantum fluctuations at order $ 1/N $ at the $ U(1) $ limit $ g^{\prime}=0 $ and near the
    $ U(1) $ limit $ g^{\prime}/g = \beta $ not too far away from the QCP.
    It is these  quantum fluctuations which lead to highly interesting quantum phenomena.
    For example, they lift the flat zero ( Goldstone )  mode at $ N=\infty $ to the oscillating shape
    shown in Fig.3a with the corresponding spectral weights shown in Fig. 3b in \cite{gold}.
    The crucial Berry phase effects only show up at a finite $ N $.
    It is the Berry phase which leads to the oscillating shape of the Goldstone mode shown in Fig.3a in \cite{gold}.
    All these important quantum effects get quenched in the  $ N= \infty $ ( classical ) limit.
    For example, the Goldstone mode is quenched to the flat zero mode \cite{gold,berryphase}.
    The amplitude mode  shown in Fig.5b in \cite{comment} is nothing but the the Higgs mode stressed in the title of
    and also fully discussed in \cite{gold}.
      We also computed the $ 1/N $ quantum fluctuations to the Higgs mode shown
      in Fig.5a and it spectral weight in Fig.5b \cite{gold}.
      Our $ 1/N $ quantum fluctuation results in Fig.3-4 match nearly perfectly well with the Exact diagonization (ED) results
      for $ N $ gets as small as $ N=2 $.


      In a recent preprint \cite{u1z2beta}, using Eqn.12,
      we investigated the quantum fluctuations at order $ 1/N $ at any $ 0 \leq g^{\prime}/g=\beta \leq 1 $ and full interaction strength.



 {\bf 3.} The last part ( about 1/10 of the paper ) of Ref.\cite{comment}  sketched a  circuit QED  implementation ( Fig.6)  of
    the $ U(1)/Z_2 $ Dicke model studied in \cite{gold}. In fact, the experimental implementations using both cold atoms inside a
    cavity and superconducting qubits inside a microwave cavity have been briefly discussed in \cite{gold,berryphase}.
    A recent experiment \cite{expggprime} realized the open version of the $ U(1)/Z_2 $
    Dicke model by using cavity-assisted Raman transitions with cold atoms inside a cavity.
    We expect that the $ U(1)/Z_2 $ Dicke model can also be realized in superconducting qubits inside a microwave cavity.
    Then all the results on Goldstone and Higss modes shown in Fig. 1-5 at and near the $ U(1) $ limit in \cite{gold}
    and  many other novel phenomena demonstrated in \cite{u1z2beta} at generic $ \beta $ can be detected  for a few $ N =2 \sim 9 $
    qubits.

    We acknowledge supports from NSF-DMR-1161497 and NSFC-11174210.


\medskip

Yu Yi-Xiang$^{1,2}$, Jinwu Ye $^{2,3}$, W.M. Liu $^{1}$ and  CunLin Zhang$^{3}$

\hspace{0.2in}\begin{minipage}{3.0in}
   $^{1}$  Beijing National Laboratory for Condensed Matter Physics, Institute of Physics, Chinese Academy of
   Sciences, Beijing 100190, China \\
   $^{2}$  Department of Physics and Astronomy, Mississippi State  University, P. O. Box 5167, Mississippi State, MS, 39762   \\
   $^{3}$ Key Laboratory of Terahertz Optoelectronics, Ministry of Education, Department of Physics, Capital Normal University, Beijing 100048, China
\end{minipage}



\end{document}